\documentclass[11pt]{article}

\usepackage{arxiv}
\usepackage{amsmath,amssymb}
\usepackage{natbib}
\usepackage{url}
\usepackage{fancyvrb}
\usepackage[T1]{fontenc}
\usepackage[utf8]{inputenc}
\usepackage{microtype}

\newcommand{\pkg}[1]{\textbf{#1}}
\newcommand{\code}[1]{\texttt{#1}}
\newcommand{\R}{\textsf{R}}

\newenvironment{Schunk}{\par\smallskip}{\par\smallskip}
\DefineVerbatimEnvironment{Sinput}{Verbatim}{fontsize=\small}
\DefineVerbatimEnvironment{Soutput}{Verbatim}{fontsize=\small}

\title{\pkg{SSLfmm}: An R Package for Semi-Supervised Learning with Mixed Missingness}

\author{
  Geoffrey J. McLachlan \\
  School of Mathematics and Physics \\
  The University of Queensland \\
  St Lucia, Queensland 4072, Australia \\
  \texttt{g.mclachlan@uq.edu.au}
  \And
  Jinran Wu \\
  School of Mathematics and Physics \\
  The University of Queensland \\
  St Lucia, Queensland 4072, Australia \\
  \texttt{jinran.wu@uq.edu.au}
}

\date{}
\renewcommand{\headeright}{McLachlan and Wu}
\renewcommand{\shorttitle}{\pkg{SSLfmm}: Semi-Supervised Learning with Mixed Missingness}

\begin{document}
\maketitle

\begin{abstract}
Partially labelled samples arise when features are observed for data, but class labels are available for only a subset. In such settings, the mechanism governing label availability may itself contain information relevant to classification, yet it is typically left unmodelled in standard semi-supervised learning procedures. The \pkg{SSLfmm} package implements likelihood-based Gaussian finite-mixture classification in which the label-missingness process is modelled jointly with the class distribution. It supports complete-case, missing completely at random (MCAR), entropy-based missing at random (MAR), and mixed analyses in which MCAR and MAR mechanisms may both contribute. For the mixed mechanism, the source of a missing label may be known or unknown. A common \R{} interface is provided for model fitting, prediction, performance assessment, and simulation. We describe the statistical formulation and software implementation and demonstrate its use through reproducible simulation and a semi-synthetic Blood Transfusion application.
\end{abstract}

\section{Introduction}\label{sec:introduction}

Semi-supervised learning (SSL) concerns statistical learning when class labels are available for only part of a training sample. In the setting considered here, a feature vector is observed for every training observation, whereas the associated class label is recorded only for a subset. Such partially classified samples arise whenever measurements can be obtained routinely, but reliable labels require additional effort, expert adjudication, laboratory confirmation, or follow-up. The statistical problem therefore lies between supervised classification, where all training labels are observed, and unsupervised clustering, where none are observed. Early likelihood-based treatments of partially classified samples were developed in discriminant analysis by~\citet{mclachlan1975iterative}, \citet{ganesalingam1978efficiency}, and \citet{oneill1978normal}. Modern SSL now includes generative, discriminative, graph-based, margin-based, self-training, and representation-learning approaches; broad statistical and machine-learning perspectives are given by~\citet{vanengelen2020survey} and~\citet{ahfock2023semi}.

Finite mixture models provide a particularly natural probabilistic framework for partially labelled data because clustering, classification, density estimation, and semi-supervised learning can be expressed through the same latent class structure. In a model-based formulation, each class is associated with a mixture component, labelled observations contribute component-specific densities, and unlabelled observations contribute through the mixture density and posterior class probabilities. This makes it possible to use both labelled and unlabelled observations within a likelihood and to retain probabilistic measures of classification uncertainty. Gaussian mixtures remain especially important because of their interpretability, their connection with classical discriminant analysis, and the large body of computational methodology based on the EM family of algorithms; a broad review of finite mixture models is given by~\citet{mclachlan2019finite}. At the same time, the software ecosystem for mixture modelling has developed far beyond a single Gaussian-mixture implementation, spanning parsimonious covariance structures, high-dimensional models, non-Gaussian and robust component distributions, Bayesian mixtures, functional and ordinal data, and incomplete observations~\citep{tong2025mixturemissing}.

A separate issue, central to the present work, is the process by which labels become unavailable. An unobserved class label is not merely an unknown latent class; it can also be the outcome of a data-collection mechanism. Under missing completely at random (MCAR), label availability is unrelated to the observed feature vector and its class. Under missing at random (MAR), the probability that a label is unavailable may depend on observed features but not on the missing class label itself~\citep{rubin1976inference,mealli2015clarifying}. In classification problems, this dependence can be scientifically plausible because observations near a decision boundary, or observations with diffuse posterior class probabilities, may be harder to label than observations that lie well inside a class. Entropy, posterior probabilities, classification margins, and related uncertainty measures have therefore been used to model informative label availability \citep{ahfock2020apparent,sportisse2023labels,lyu2024semi,wu2026informative,liao2026margin}. When the class model parameters also enter the missing-label mechanism, the observation process carries information about the classifier and should be represented explicitly in the likelihood.

We proceed here with maximum-likelihood estimation of the Bayes classification rule under three categories of label missingness: MCAR, MAR, and MCAR/MAR. Note that the notation MCAR/MAR means MCAR or MAR, or both. In this paper, we refer to the third category simply as the \emph{mixed} mechanism. Under MCAR, maximum likelihood can be implemented directly with the EM algorithm: labelled observations contribute through their class-specific normal densities, whereas unlabelled observations contribute through the corresponding finite-mixture density. Software for this type of mixture fitting was surveyed by~\citet{mclachlan2000finite}, with more recent developments reviewed by~\citet{mclachlan2019finite}; additional implementations relevant to the present setting are summarised below. For MAR mechanisms in which label missingness is related to classification entropy, likelihood-based approaches were developed by~\citet{ahfock2020apparent}, with an \R{} implementation subsequently provided by~\citet{lyu2024semi}. The third, mixed category is the focus of the present software contribution.

The \pkg{SSLfmm} package was developed at this intersection of model-based SSL and missing-data modelling. It fits Gaussian finite-mixture classifiers using complete-case analysis or models based on MCAR, entropy-based MAR, and mixed label missingness. In the mixed formulation, a missing label may arise under either an MCAR mechanism or an entropy-based MAR mechanism. The missingness source may be known, for example, when a study records why a label was not obtained, or unknown, so that only the aggregate missing-label indicator is available. These two data structures lead to related but distinct expectation--conditional maximisation (ECM) procedures \citep{meng1993ecm,mclachlankrishnan2008em}. Here we provide the formulation and ECM
estimation needed to make the software implementation self-contained,
and focus on the R interface and reproducible workflows.

\subsection{Software landscape and the role of SSLfmm}

The R ecosystem for finite mixtures is extensive, and the position of \pkg{SSLfmm} is clearest when existing software is viewed in terms of the statistical problem it is designed to solve. At the general-purpose end, \pkg{mclust} is one of the most established implementations for Gaussian model-based clustering, classification, density estimation, and discriminant analysis, with parsimonious covariance parametrisation and semi-supervised classification available through \code{MclustSSC}() \citep{fraley2007model,scrucca2016mclust,mclust2026}. The \pkg{Rmixmod} package provides an R interface to the MIXMOD library for unsupervised, supervised, and semi-supervised mixture classification and supports a broad family of covariance structures and model-selection criteria \citep{lebret2015rmixmod}. The \pkg{mixture} package offers Gaussian-mixture clustering and classification \citep{pocuca2021mixture}, whereas \pkg{mixtools} provides a collection of parametric, semiparametric, and regression-mixture procedures \citep{benaglia2010mixtools}. More general mixture infrastructure is supplied by \pkg{flexmix}, which separates the mixture machinery from model-specific component fitting and supports, among other models, mixtures of regressions and generalised linear models \citep{grun2008flexmix}. Together, these packages demonstrate the maturity of finite-mixture computation in R, but their principal modelling target is the component distribution and latent membership structure rather than a stochastic model for label availability.

Earlier software developments include the \pkg{EMMIX} suite for fitting normal and $t$-mixture models \citep{mclachlan1999emmix}. Later EMMIX-related implementations include \pkg{EMMIXmfa} and \pkg{EMMIXmcfa} for factor-analytic mixture models, \pkg{mmtfa} for mixtures of $t$-factor analyzers, and \pkg{EMMIXskew} and \pkg{EMMIXcskew} for skew-normal and skew-$t$ mixture models \citep{emmixmfa2024,emmixmcfa2017,mmtfa2018,emmixskew2018,lee2018emmixcskew}. The \pkg{mixsmsn} package provides related functionality for mixtures based on scale mixtures of skew-normal distributions \citep{prates2013mixsmsn}. These developments are reviewed in broader context by~\citet{mclachlan2019finite}.

A substantial branch of the software literature addresses high-dimensional or non-standard data structures. The \pkg{pgmm} package implements parsimonious Gaussian mixture models for high-dimensional clustering through latent factor structures \citep{mcnicholas2008pgmm,mcnicholas2021pgmm}, while \pkg{HDclassif} provides model-based clustering and discriminant analysis in high dimension using parsimonious Gaussian models in discriminative subspaces \citep{berge2012hdclassif}. Mixture methods have also been adapted to data types for which ordinary multivariate Gaussian components are inappropriate. The \pkg{BayesBinMix} package provides Bayesian model-based clustering for multivariate binary data \citep{papastamoulis2017bayesbinmix}; \pkg{ordinalClust} targets clustering, co-clustering, and classification of ordinal observations \citep{selosse2021ordinalclust}; and \pkg{funFEM} performs model-based clustering of functional data in a discriminative functional subspace \citep{bouveyron2021funfem}. These packages illustrate a general pattern in statistical software: a mixture model becomes a distinct software contribution when the implementation supports a specific data structure, inferential target, or practical workflow that is not covered by general mixture engines.

Another major line of development replaces Gaussian components with robust or more flexible distributions. The \pkg{teigen} package implements model-based clustering and classification with mixtures of multivariate $t$ distributions \citep{andrews2018teigen}. The \pkg{ContaminatedMixt} package fits parsimonious mixtures of contaminated-normal distributions and provides automatic detection of mild outliers \citep{punzo2018contaminated}. The \pkg{MixGHD} package extends model-based clustering, classification, and discriminant analysis to generalised hyperbolic distributions, allowing component-specific skewness and heavier tails \citep{tortora2021mixghd}. A complementary robust strategy is implemented in \pkg{tclust}, which uses trimming to reduce the effect of gross outliers \citep{fritz2012tclust}. These developments broaden the class of feature distributions that can be represented. In contrast, \pkg{SSLfmm} keeps a Gaussian component model and changes a different part of the data-generating process: the mechanism governing whether a class label is observed.

Bayesian mixture software forms another mature branch. The \pkg{REBayes} package implements empirical-Bayes mixture methods using convex optimisation \citep{koenker2017rebayes}; \pkg{bayesmix} provides Bayesian mixture modelling through \pkg{JAGS} \citep{grun2023bayesmix}; and \pkg{BNPmix} provides Bayesian nonparametric density estimation, clustering, and regression using Pitman--Yor mixtures \citep{corradin2021bnpmix}. These packages are useful when uncertainty about the mixing distribution, component structure, or number of clusters is the primary target. The emphasis of \pkg{SSLfmm} is different: it deliberately retains a finite Gaussian-mixture classifier so that the class-label observation mechanism, and in particular the distinction between the MCAR and entropy-based MAR mechanisms, can be modelled and interpreted directly.

Missing-data software intersects mixture modelling in a different way. Most available tools address missing feature values, either by preprocessing or within the model-fitting algorithm. The package \pkg{mix} provides likelihood-based estimation and multiple imputation for mixed categorical and continuous data \citep{schafer2017mix}. The \pkg{ClustImpute} package combines imputation with $k$-means clustering \citep{pfaffel2020clustimpute}, while the methodology underlying \pkg{miclust} combines multiple imputation with cluster analysis \citep{basagana2013miclust}. The \pkg{VarSelLCM} package performs variable selection in model-based clustering of mixed data with missing values \citep{marbac2019varsel}. More directly, mixture-based approaches include \pkg{MGMM}, which provides missingness-aware Gaussian mixture models \citep{mccaw2023mgmm}, and \pkg{RMixtComp}, which fits mixture models for heterogeneous and partially missing data \citep{kubicki2023rmixtcomp}. The \pkg{mixture} package also incorporates missing feature values within Gaussian-mixture clustering and classification, and \pkg{mclust} provides imputation facilities as part of its broader model-based workflow. These tools solve important incomplete-data problems, but the incomplete object is the feature vector rather than the class label.

The distinction is especially clear in the recent \pkg{MixtureMissing} package. \pkg{MixtureMissing} directly accounts for incomplete feature vectors in model-based clustering and supports a wide family of normal, contaminated-normal, generalised-hyperbolic, skew-$t$, and related distributions \citep{tong2025mixturemissing}. Its scope reflects a substantial literature on robust and flexible clustering with incomplete measurements. \pkg{SSLfmm} addresses the complementary problem in which the feature vector is fully observed, but the class label is only partially observed. Thus, although both packages are concerned with missingness in finite-mixture models, they model different missing-data objects and answer different scientific questions.

The semi-supervised learning software landscape provides the closest comparison. Both \pkg{mclust} and \pkg{Rmixmod} can combine labelled and unlabelled observations within mixture models. The \pkg{RSSL} package provides a broader collection of semi-supervised classifiers, including implicitly constrained procedures, self-learning methods, transductive support vector machines, and graph- or manifold-based approaches \citep{krijthe2016rssl}. These tools primarily exploit unlabelled observations for prediction. They generally do not parameterise the process by which labels are missing. The closest existing software to \pkg{SSLfmm} is \pkg{gmmsslm}, which implements semi-supervised Gaussian mixtures with an entropy-based missing-label mechanism \citep{lyu2024semi}. That work demonstrates that the pattern of missing labels can itself be informative when missingness depends on classification uncertainty. Table~\ref{tab:software_comparison} summarises the main distinctions among selected packages that are most relevant to partially labelled mixture-model classification.

\begin{table}[!ht]
\centering
\small
\begin{tabular}{lccccc}
\hline
Package & Semi-supervised & Missing & Explicit label & Mixed & Unknown \\
&classification & features & missingness & mechanism & missingness \\
&&&mechanism &&source \\
\hline
\pkg{mclust}          & Yes & Imputation & No  & No  & No \\
\pkg{Rmixmod}         & Yes & --         & No  & No  & No \\
\pkg{RSSL}            & Yes & --         & No  & No  & No \\
\pkg{MixtureMissing}  & No  & Yes        & No  & No  & No \\
\pkg{gmmsslm}         & Yes & No         & Yes & No  & No \\
\textbf{\pkg{SSLfmm}} & Yes & No         & Yes & Yes & Yes \\
\hline
\end{tabular}
\caption{Comparison of selected \R{} packages relevant to partially labelled mixture-model classification. For \pkg{mclust}, ``Imputation'' indicates that imputation facilities are available within its broader model-based workflow rather than through a dedicated missing-feature model. \pkg{MixtureMissing} focuses on model-based clustering with incomplete feature vectors rather than partially labelled semi-supervised classification.}
\label{tab:software_comparison}
\end{table}

Seen against this broader software landscape, the contribution of \pkg{SSLfmm} is not to provide another generic clustering engine or another catalogue of mixture distributions. Instead, the package connects two established software traditions that have largely developed separately: model-based semi-supervised classification and likelihood-based modelling of missing data. Its distinguishing data structure is a partially classified sample in which label absence can arise under either an MCAR mechanism or an entropy-based MAR mechanism. Moreover, the missingness source may be known or unknown. The package therefore supports complete-case, MCAR, MAR, and mixed analyses within a common interface; provides distinct ECM algorithms for known and unknown missingness sources; returns fitted objects supporting printing, summaries, and prediction; supplies simulation functions that preserve the missingness-source information; and provides tools for classification-performance assessment. This design makes it possible to study not only how unlabelled observations contribute to classification, but also how assumptions about the generation of those unlabelled observations affect fitted mixture parameters, estimated missingness parameters, and predictive uncertainty.

The remainder of the article is organised as follows. Section 2 formulates the mixed label-missingness model and develops the ECM algorithms for known and unknown missingness sources. Section 3 describes the main user-facing functions and the common fitted-object interface. Section 4 presents a simulation illustration comparing the known- and unknown-source formulations. Section 5 illustrates the broader modelling workflow through a semi-synthetic application to the Blood Transfusion data, comparing MCAR, MAR, and mixed analyses under two covariance specifications. Section 6 concludes with final remarks.

\section{Problem formulation and ECM estimation}\label{sec:problem} 
 
\subsection{Partially classified Gaussian mixtures} 
 
Let $\boldsymbol{Y}_j$ be the $p$-dimensional feature vector for observation $j$ and let $Z_j\in\{1,\ldots,g\}$ be its class label. We write $\boldsymbol{y}_j$ and $z_j$ for their realised values. Conditional on $Z_j=i$, the feature vector is assumed to have a multivariate Gaussian density 
\begin{equation*} 
  f_i(\boldsymbol{y};\boldsymbol{\omega}_i)=\phi_p(\boldsymbol{y};\boldsymbol{\mu}_i,\boldsymbol{\Sigma}_i), 
  \qquad i=1,\ldots,g, 
  \end{equation*} 
with prior class probabilities $\pi_1,\ldots,\pi_g$, where $\sum_i\pi_i=1$. We use $\boldsymbol{\theta}$ to collect the mixing proportions and component parameters $\{\boldsymbol{\omega}_i\}_{i=1}^g$. The marginal density is 
\begin{equation*} 
  f_{\boldsymbol{Y}}(\boldsymbol{y};\boldsymbol{\theta})
  =\sum_{i=1}^g \pi_i f_i(\boldsymbol{y};\boldsymbol{\omega}_i), 
  \end{equation*} 
and the posterior probability that $\boldsymbol{y}$ belongs to class $i$ is 
\begin{equation*} 
  \tau_i(\boldsymbol{y};\boldsymbol{\theta})= 
  \frac{\pi_i f_i(\boldsymbol{y};\boldsymbol{\omega}_i)}
       {\sum_{h=1}^g \pi_h f_h(\boldsymbol{y};\boldsymbol{\omega}_h)}. 
  \end{equation*} 
The Bayes classifier allocates $\boldsymbol{y}$ to the class with the largest posterior probability. 
 
Classification uncertainty is measured by Shannon entropy, 
\begin{equation*} 
  e(\boldsymbol{y};\boldsymbol{\theta})
  =-\sum_{i=1}^g \tau_i(\boldsymbol{y};\boldsymbol{\theta})
    \log \tau_i(\boldsymbol{y};\boldsymbol{\theta}). 
  \end{equation*} 

Let $M_{1j}$ and $M_{2j}$ denote the missing-label indicators for the MCAR and entropy-based MAR mechanisms, respectively. These indicators are mutually exclusive, and the aggregate missing-label indicator is
\begin{equation*}
  M_j=M_{1j}+M_{2j}.
  \end{equation*}
Thus, $M_j=0$ denotes a labelled observation, whereas $M_j=1$ denotes an observation whose label is missing under either the MCAR or the entropy-based MAR mechanism.
 
For the entropy-based MAR mechanism, define 
\begin{equation} 
  q_j=q(\boldsymbol{y}_j;\boldsymbol{\theta},\boldsymbol{\xi}) 
  =\operatorname{logit}^{-1}
  \left\{\xi_0+\xi_1\log e(\boldsymbol{y}_j;\boldsymbol{\theta})\right\}, 
  \qquad \xi_1>0. 
  \label{eq:q} 
\end{equation} 
With $0\leq\alpha<1$ denoting the MCAR probability, the conditional probabilities of the three mutually exclusive missingness states are
\begin{align*}
  \Pr(M_{1j}=1,M_{2j}=0\mid \boldsymbol{y}_j)
    &= \alpha, \\
  \Pr(M_{1j}=0,M_{2j}=1\mid \boldsymbol{y}_j)
    &= (1-\alpha)q_j, \\
  \Pr(M_{1j}=0,M_{2j}=0\mid \boldsymbol{y}_j)
    &= (1-\alpha)(1-q_j).
\end{align*}

If only the aggregate indicator $M_j$ is observed, the conditional probability of a missing label is therefore 
\begin{equation*} 
  r_\alpha(\boldsymbol{y}_j;\boldsymbol{\theta},\boldsymbol{\xi}) 
  =\Pr(M_j=1\mid \boldsymbol{y}_j) 
  =\alpha+(1-\alpha)q_j. 
  \end{equation*} 
Because the entropy depends on the mixture parameters through the posterior probabilities, the missing-label mechanism contributes to estimation of the mixture model and is retained in the likelihood. 
 
\subsection{A common E-step for the missing class labels} 
 
Write $z_{ij}=1$ if observation $j$ belongs to class $i$ and zero otherwise, and let $m_j$ denote the observed realisation of $M_j$. At ECM iteration $k$, the posterior class probabilities are 
\begin{equation*} 
  \tau_{ij}^{(k)}= 
  \frac{\pi_i^{(k)}f_i(\boldsymbol{y}_j;\boldsymbol{\omega}_i^{(k)})} 
       {\sum_{h=1}^g\pi_h^{(k)}f_h(\boldsymbol{y}_j;\boldsymbol{\omega}_h^{(k)})}. 
  \end{equation*} 
For a missing class label, its complete-data indicator is replaced by its conditional expectation. Hence 
\begin{equation*} 
  \widetilde z_{ij}^{(k)}
  =(1-m_j)z_{ij}+m_j\tau_{ij}^{(k)}. 
  \end{equation*} 
The entropy and log-entropy covariates at the current iterate are 
\begin{equation*} 
  e_j^{(k)}
  =-\sum_{i=1}^g \tau_{ij}^{(k)}\log\tau_{ij}^{(k)}, 
  \qquad 
  v_j^{(k)}=\log e_j^{(k)}. 
  \end{equation*} 
The expected class-model contribution common to both ECM implementations is 
\begin{equation*} 
  Q_{\mathrm{cls}}^{(k)}(\boldsymbol{\theta}) 
  =\sum_{j=1}^n\sum_{i=1}^g 
  \widetilde z_{ij}^{(k)}
  \{\log\pi_i+\log f_i(\boldsymbol{y}_j;\boldsymbol{\omega}_i)\}. 
  \end{equation*} 
 
\subsection{ECM with a known missingness source} 
 
Suppose that the mechanism-specific indicators $M_{1j}$ and $M_{2j}$ are known, and let $m_{1j}$ and $m_{2j}$ denote their realised values. Conditional on the class-label E-step, the expected complete-data log likelihood is 
\begin{align*} 
Q^{(\mathrm{known})}(\boldsymbol{\Psi}\mid\boldsymbol{\Psi}^{(k)}) 
={}&Q_{\mathrm{cls}}^{(k)}(\boldsymbol{\theta}) 
+\sum_{j=1}^n\Big[ 
 m_{1j}\log\alpha+(1-m_{1j})\log(1-\alpha) \\ 
&\qquad +m_{2j}\log q(\boldsymbol{y}_j;\boldsymbol{\theta},\boldsymbol{\xi}) 
 +(1-m_{1j}-m_{2j})
 \log\{1-q(\boldsymbol{y}_j;\boldsymbol{\theta},\boldsymbol{\xi})\}\Big], 
\end{align*} 
where $\boldsymbol{\Psi}=(\boldsymbol{\theta}^\top,\boldsymbol{\xi}^\top,\alpha)^\top$. 
 
The ECM cycle has three conditional maximisations. First, if $\alpha$ is not fixed by design, it has the closed-form update 
\begin{equation} 
  \widehat\alpha
  =\frac{1}{n}\sum_{j=1}^n m_{1j}. 
  \label{eq:alphaobs} 
\end{equation} 
Second, with $\boldsymbol{\theta}$ fixed at its current value, $\boldsymbol{\xi}$ is updated by maximising 
\begin{equation} 
  Q_{\boldsymbol{\xi}}^{(\mathrm{known})}(\boldsymbol{\xi}) 
  =\sum_{j=1}^n\left[ 
    m_{2j}\log q_j
    +(1-m_{1j}-m_{2j})\log(1-q_j) 
  \right], 
  \label{eq:xiobs} 
\end{equation} 
subject to $\xi_1>0$, where 
\[
q_j=\operatorname{logit}^{-1}
\{\xi_0+\xi_1v_j^{(k)}\}.
\]
Third, $\boldsymbol{\theta}$ is updated by maximising the class-model contribution together with the entropy-based missingness terms, 
\begin{align} 
  Q_{\boldsymbol{\theta}}^{(\mathrm{known})}(\boldsymbol{\theta}) 
  ={}&Q_{\mathrm{cls}}^{(k)}(\boldsymbol{\theta}) 
  +\sum_{j=1}^n m_{2j}
  \log q(\boldsymbol{y}_j;\boldsymbol{\theta},\boldsymbol{\xi}^{(k+1)}) \notag\\ 
  &+\sum_{j=1}^n(1-m_{1j}-m_{2j}) 
  \log\{1-q(\boldsymbol{y}_j;\boldsymbol{\theta},\boldsymbol{\xi}^{(k+1)})\}. 
  \label{eq:thetaobs} 
\end{align} 
The entropy is recomputed as $\boldsymbol{\theta}$ varies in this conditional maximisation because $q(\boldsymbol{y};\boldsymbol{\theta},\boldsymbol{\xi})$ depends on the posterior class probabilities. The cycle is repeated until the full log-likelihood satisfies the convergence criterion. When a numerical conditional step produces an ascent step rather than an exact conditional maximiser, the iteration is a generalised ECM step. 
 
For reference, the known-source algorithm can be summarised as follows. 
\begin{enumerate} 
  \item Initialise $\boldsymbol{\theta}^{(0)}$ and $\boldsymbol{\xi}^{(0)}$; if $\alpha$ is not fixed, compute it using Equation~\ref{eq:alphaobs}. 
  \item At iteration $k$, compute $\tau_{ij}^{(k)}$, $\widetilde z_{ij}^{(k)}$, $e_j^{(k)}$, and $v_j^{(k)}$. 
  \item Update $\boldsymbol{\xi}$ by maximising Equation~\ref{eq:xiobs} subject to $\xi_1>0$. 
  \item Update $\boldsymbol{\theta}$ by maximising Equation~\ref{eq:thetaobs}, recomputing the entropy as $\boldsymbol{\theta}$ changes. 
  \item Evaluate the full log-likelihood and iterate until convergence. 
\end{enumerate} 
 
\subsection{ECM with an unknown missingness source} 
 
When only the aggregate indicator $M_j$ is available, the mechanism-specific indicators $M_{1j}$ and $M_{2j}$ are unknown for observations with missing labels. For an observation with $m_j=1$, define 
\begin{equation} 
  u_{1j}^{(k)} 
  =\mathbb{E}(M_{1j}\mid m_j=1,\boldsymbol{y}_j;\boldsymbol{\Psi}^{(k)}) 
  =\frac{\alpha^{(k)}} 
  {\alpha^{(k)}+(1-\alpha^{(k)})q_j^{(k)}}, 
  \label{eq:u1} 
\end{equation} 
and
\begin{equation} 
  u_{2j}^{(k)} 
  =\mathbb{E}(M_{2j}\mid m_j=1,\boldsymbol{y}_j;\boldsymbol{\Psi}^{(k)}) 
  =\frac{(1-\alpha^{(k)})q_j^{(k)}} 
  {\alpha^{(k)}+(1-\alpha^{(k)})q_j^{(k)}}. 
  \label{eq:u2} 
\end{equation} 
For $m_j=0$, set $u_{1j}^{(k)}=u_{2j}^{(k)}=0$. The expected complete-data criterion becomes 
\begin{align*} 
Q^{(\mathrm{unknown})}(\boldsymbol{\Psi};\boldsymbol{\Psi}^{(k)}) 
={}&Q_{\mathrm{cls}}^{(k)}(\boldsymbol{\theta}) 
+\sum_{j=1}^n\Big[ 
 u_{1j}^{(k)}\log\alpha
 +(1-u_{1j}^{(k)})\log(1-\alpha) \\ 
&\qquad 
 +u_{2j}^{(k)}\log q(\boldsymbol{y}_j;\boldsymbol{\theta},\boldsymbol{\xi}) 
 +\{1-u_{1j}^{(k)}-u_{2j}^{(k)}\} 
 \log\{1-q(\boldsymbol{y}_j;\boldsymbol{\theta},\boldsymbol{\xi})\}\Big]. 
\end{align*} 
The conditional update for the MCAR probability is 
\begin{equation} 
  \alpha^{(k+1)}
  =\frac{1}{n}\sum_{j=1}^n u_{1j}^{(k)}. 
  \label{eq:alphalat} 
\end{equation} 
Given $\boldsymbol{\theta}^{(k)}$, the update for $\boldsymbol{\xi}$ maximises 
\begin{equation} 
  Q_{\boldsymbol{\xi}}^{(\mathrm{unknown})}(\boldsymbol{\xi}) 
  =\sum_{j=1}^n\left[ 
    u_{2j}^{(k)}\log q_j 
    +\{1-u_{1j}^{(k)}-u_{2j}^{(k)}\}\log(1-q_j) 
  \right], 
  \label{eq:xilat} 
\end{equation} 
again subject to $\xi_1>0$. Finally, $\boldsymbol{\theta}$ is updated by maximising 
\begin{align} 
  Q_{\boldsymbol{\theta}}^{(\mathrm{unknown})}(\boldsymbol{\theta}) 
  ={}&Q_{\mathrm{cls}}^{(k)}(\boldsymbol{\theta}) 
  +\sum_{j=1}^n u_{2j}^{(k)} 
  \log q(\boldsymbol{y}_j;\boldsymbol{\theta},\boldsymbol{\xi}^{(k+1)}) \notag\\ 
  &+\sum_{j=1}^n\{1-u_{1j}^{(k)}-u_{2j}^{(k)}\} 
  \log\{1-q(\boldsymbol{y}_j;\boldsymbol{\theta},\boldsymbol{\xi}^{(k+1)})\}. 
  \label{eq:thetalat} 
\end{align} 
 
Because the unknown-source likelihood can have relatively flat regions in directions that separate the two missingness mechanisms, multiple initial values are useful. The unknown-source algorithm is therefore summarised as follows. 
\begin{enumerate} 
  \item Choose several initial values for $\boldsymbol{\theta}^{(0)}$, $\boldsymbol{\xi}^{(0)}$, and $\alpha^{(0)}$. 
  \item Compute the class posterior probabilities, the class-label expectations, and the log-entropy covariates. 
  \item For observations with missing labels, compute $u_{1j}^{(k)}$ and $u_{2j}^{(k)}$ from Equations~\ref{eq:u1}--\ref{eq:u2}. 
  \item Update $\alpha$ using Equation~\ref{eq:alphalat}. 
  \item Update $\boldsymbol{\xi}$ by maximising Equation~\ref{eq:xilat} subject to $\xi_1>0$. 
  \item Update $\boldsymbol{\theta}$ by maximising Equation~\ref{eq:thetalat}, recomputing the entropy during the conditional maximisation. 
  \item For each starting value, repeat Steps 2--6 until convergence and evaluate the marginal observed-data log-likelihood. Among the converged runs, retain the solution with the largest log-likelihood. 
\end{enumerate} 
 
The two algorithms differ only in whether the mechanism-specific missingness indicators are known or, when unknown, replaced in the E-step by their conditional expectations. This gives a common statistical model but two practically distinct data-collection settings. Entropy-based MAR and mixed fitting require $g \geq 2$, since posterior
entropy is constant when $g=1$ and the entropy slope is therefore not
identifiable.

\section{The \pkg{SSLfmm} package}\label{sec:package}

The package \pkg{SSLfmm} \citep{SSLfmm2026} requires \R{} version 3.6.0 or later \citep{R2026}, imports only the standard \pkg{graphics} and \pkg{stats} packages, and is distributed under the GPL-3 license. The \pkg{SSLfmm} package is available from the Comprehensive R Archive Network (CRAN). Development is version controlled at \url{https://github.com/wujrtudou/SSLfmm}, which also provides the public issue tracker. The package can be installed with

\begin{Schunk}
\begin{Sinput}
R> install.packages("SSLfmm")
\end{Sinput}
\end{Schunk}

The package is organised around a small number of user-facing functions. A standard analysis can therefore be performed without calling likelihood, parameter-packing, or optimisation helpers directly. The principal functions are described below.

\begin{description}
  \item[\code{fit\_sslfmm}()] fits Gaussian-mixture classifiers using complete-case, MCAR, entropy-based MAR, or mixed analyses using the estimation procedures described in Section~\ref{sec:problem}. The argument \code{method} selects the missing-label model, while \code{indicator = "known"} or \code{"unknown"} specifies whether the missingness source is known or unknown in the mixed model. The number of mixture components is supplied through \code{g}; common or component-specific covariance matrices are selected through \code{covariance\_type}.

  \item[\code{initialize\_sslfmm}()] constructs starting values for the mixture model. Observed labels anchor component identities, while unlabelled observations can contribute to initialisation when needed. The returned initialisation object can be inspected before fitting.

  \item[\code{simulate\_sslfmm}()] generates Gaussian-mixture samples with no missing labels or with MCAR, entropy-based MAR, or mixed label missingness. It is useful for checking an analysis workflow and for reproducible examples.

  \item[\code{simulate\_mixed\_missingness}()] is a convenience interface for the mixed mechanism. Its returned object contains the observed partially labelled data, the generating parameter values, mechanism-specific groups, the MAR probabilities, and the underlying complete simulated sample.

  \item[\code{rmix}()] generates observations from a Gaussian finite mixture and is the low-level simulation function used by the higher-level simulation routines.

  \item[\code{classification\_performance}()] evaluates predicted classes against known class labels. It returns accuracy, error rate, balanced accuracy, macro precision, macro recall, and macro F1; when posterior probabilities are supplied, it also reports log loss. Macro averages are taken over classes represented in the reference labels, and a represented class that receives no predictions contributes zero precision and zero F1 rather than being omitted from the macro average.
\end{description}

The implementation uses bounded numerical optimisation for the maximisation steps associated with the likelihood formulations in Section~\ref{sec:problem}. Multiple starting values can be used, and the solution with the largest attained log-likelihood among successful starts is retained. Covariance matrices are parameterised to maintain positive definiteness during fitting, while log-scale calculations and an entropy floor are used where appropriate for numerical stability. Multiple initial values for the MCAR contribution can additionally be supplied for the unknown-source mixed formulation.

The fitting interface follows standard \R{} conventions. The features are passed as an $n\times p$ numeric matrix or data frame \code{x}, and the class labels are passed through \code{y}, with \code{NA} for unlabelled observations. Numeric, character, and factor labels can be used. In a known-source mixed analysis, \code{missing\_source} has one entry per observation and identifies missing-label rows as \code{"mcar"} or \code{"mar"}; a logical or 0--1 MCAR-source indicator is also accepted. Values supplied on labelled rows are ignored and internally treated as \code{"observed"}. For an unknown-source fit, this argument is omitted.

A successful call to \code{fit\_sslfmm}() returns an object of class \pkg{SSLfmm}. The object stores the fitted mixing proportions, component means and covariance matrices, posterior probabilities, entropy values, fitted missingness parameters, maximised log-likelihood, optimiser and convergence information, and per-start summaries. For an unknown-source mixed fit, it also stores the fitted posterior probability that each missing label arose through the MCAR channel. The fitted object can be inspected and used for prediction through the following methods:

\begin{Schunk}
\begin{Sinput}
R> print(fit)
R> summary(fit)
R> predict(fit, newdata = xnew, type = "all")
\end{Sinput}
\end{Schunk}

The \code{predict}() method can return predicted classes, posterior probabilities, entropy, or all three. The same methods are used for known- and unknown-source fits, so downstream analysis does not depend on how the missingness source was handled during estimation.

Input checking is performed before fitting. In particular, the package checks the dimensions and finiteness of the feature matrix, the consistency of the observed labels with the requested number of components, the covariance specification, and the availability of both labelled and unlabelled observations when required by the selected method. For mixed models with a known source, the \code{missing\_source} vector is also validated. These checks are useful in applications because errors in the coding of partially observed labels can otherwise be difficult to distinguish from numerical fitting problems.

\section{A reproducible simulation illustration}\label{sec:example}

This example compares the mixed model when the missingness source is known with the corresponding model in which the source is unknown. The same simulated sample is used for both fits. The simulation is chosen to provide a useful range of classification uncertainty. In particular, the two Gaussian classes overlap sufficiently that the entropy covariate is not concentrated near zero. A sample size of 1000 and multiple optimisation starts are used to estimate the unknown-source missingness parameters more stably.

\begin{Schunk}
\begin{Sinput}
R> library("SSLfmm")
R> set.seed(123)
R> n <- 1000L
R> g <- 2L
R> pi_true <- c(0.5, 0.5)
R> mu_true <- matrix(c(-1.2, -0.8, 1.2, 0.8), nrow = 2, ncol = 2)
R> sigma_true <- matrix(c(1.0, 0.2, 0.2, 1.0), nrow = 2, byrow = TRUE)
\end{Sinput}
\end{Schunk}

Labels are removed under a mixed mechanism with MCAR probability $\alpha=0.10$. The entropy-based MAR mechanism has a target average probability of 0.25 and slope $\xi_1=1.5$.

\begin{Schunk}
\begin{Sinput}
R> sim <- simulate_mixed_missingness(
+    n = n, pi = pi_true, mu = mu_true, sigma = sigma_true,
+    alpha = 0.10, mar_rate = 0.25, xi1 = 1.5, seed = 123)
R> x <- as.matrix(sim$data[, c("x1", "x2")])
R> y <- sim$data$label
R> truth <- sim$data$truth
R> id_mis <- is.na(y)
R> mean(id_mis)
\end{Sinput}
\begin{Soutput}
[1] 0.319
\end{Soutput}
\begin{Sinput}
R> table(sim$data$missing_source)
\end{Sinput}
\begin{Soutput}
     mar     mcar observed
     217      102      681
\end{Soutput}
\begin{Sinput}
R> round(c(alpha = sim$true_setup$alpha,
+    xi0 = unname(sim$true_setup$xi["xi0"]),
+    xi1 = unname(sim$true_setup$xi["xi1"])), 3)
\end{Sinput}
\begin{Soutput}
alpha   xi0   xi1
0.100 1.385 1.500
\end{Soutput}
\end{Schunk}

Thus, 31.9\% of the class labels are unavailable during fitting. The value of $\xi_0$ is calibrated internally to attain the requested average MAR probability for the realised sample. The moderate overlap between the two classes keeps the log-entropy covariate in an informative range and avoids the near-separation that can occur when the classes are almost perfectly separated.

\subsection{Known- and unknown-source fits}

The known-source fit uses the known missingness source (MCAR or MAR). The unknown-source fit uses only the aggregate pattern of missing class labels. Both models are fitted using multiple optimisation starts.

\begin{Schunk}
\begin{Sinput}
R> fit_known <- fit_sslfmm(
+    x = x, y = y, g = g, method = "mixed", covariance_type = "equal",
+    indicator = "known", missing_source = sim$data$missing_source,
+    n_starts = 10,
+    control = list(iter.max = 1500, eval.max = 2500, rel.tol = 1e-8),
+    seed = 2024)
R> fit_unknown <- fit_sslfmm(
+    x = x, y = y, g = g, method = "mixed", covariance_type = "equal",
+    indicator = "unknown", n_starts = 10,
+    alpha_starts = c(0.02, 0.06, 0.10, 0.15, 0.25),
+    control = list(iter.max = 1500, eval.max = 2500, rel.tol = 1e-8),
+    seed = 2024)
\end{Sinput}
\end{Schunk}

The same prediction interface is used for the two fitted models.

\begin{Schunk}
\begin{Sinput}
R> pred_known <- predict(fit_known, newdata = x, type = "all")
R> pred_unknown <- predict(fit_unknown, newdata = x, type = "all")
\end{Sinput}
\end{Schunk}

\subsection{Classification performance}

Because the complete simulated labels are available, classification can be evaluated specifically on the observations whose class labels are missing during fitting. Their feature vectors are part of the fitted semi-supervised sample, so this evaluation concerns the unlabelled portion of the fitted sample rather than an independent test set.

\begin{Schunk}
\begin{Sinput}
R> perf_known <- classification_performance(
+    truth[id_mis], pred_known$class[id_mis],
+    pred_known$posterior[id_mis, , drop = FALSE])
R> perf_unknown <- classification_performance(
+    truth[id_mis], pred_unknown$class[id_mis],
+    pred_unknown$posterior[id_mis, , drop = FALSE])
R> metric_names <- c("accuracy", "balanced_accuracy", "macro_f1", "log_loss")
R> results_class <- rbind(
+    "Known source" = perf_known$metrics[metric_names],
+    "Unknown source" = perf_unknown$metrics[metric_names])
R> colnames(results_class) <- c("Accuracy", "Balanced accuracy", "Macro F1", 
+     "Log loss")
R> round(results_class, 3)
\end{Sinput}
\begin{Soutput}
               Accuracy Balanced accuracy Macro F1 Log loss
Known source       0.806             0.806    0.806    0.406
Unknown source     0.809             0.809    0.809    0.407
\end{Soutput}
\end{Schunk}

The two fits give very similar classification performance on the observations with missing labels. The known-source fit has accuracy, balanced accuracy, and macro F1 equal to 0.806, whereas the corresponding values for the unknown-source fit are 0.809. The known-source fit has a slightly smaller log loss, 0.406 compared with 0.407. These values summarise one reproducible simulated sample and are not intended as a general ranking of the two formulations.

\subsection{Missingness-parameter estimates}

The generating and fitted missingness parameters can be inspected directly. The chosen sample size and non-degenerate entropy distribution are intended to make the unknown-source decomposition of the MCAR and MAR channels more stable.

\begin{Schunk}
\begin{Sinput}
R> results_missingness <- data.frame(
+    Model = c("Truth", "Known source", "Unknown source"),
+    alpha = c(sim$true_setup$alpha, fit_known$alpha, fit_unknown$alpha),
+    xi0 = c(unname(sim$true_setup$xi["xi0"]),
+      unname(fit_known$xi["xi0"]), unname(fit_unknown$xi["xi0"])),
+    xi1 = c(unname(sim$true_setup$xi["xi1"]),
+      unname(fit_known$xi["xi1"]), unname(fit_unknown$xi["xi1"])),
+    check.names = FALSE)
R> transform(results_missingness, alpha = round(alpha, 3),
+    xi0 = round(xi0, 3), xi1 = round(xi1, 3))
\end{Sinput}
\begin{Soutput}
           Model alpha   xi0   xi1
1          Truth 0.100 1.385 1.500
2   Known source 0.102 1.443 1.674
3 Unknown source 0.105 1.533 1.788
\end{Soutput}
\end{Schunk}

The generating MCAR probability is 0.100. Its estimate is 0.102 for the known-source model and 0.105 for the unknown-source model. The fitted entropy-based MAR parameters are $(\widehat{\xi}_0,\widehat{\xi}_1)=(1.443,1.674)$ for the known-source fit and $(1.533,1.788)$ for the unknown-source fit, compared with the generating values $(1.385,1.500)$.

The simulated sample contains 1000 observations, with 31.9\% of the class labels missing. The known-source and unknown-source models are fitted to the same partially observed sample. On observations with missing labels, the known-source fit has accuracy 0.806, while the unknown-source fit has accuracy 0.809. Their balanced accuracies are 0.806 and 0.809, respectively. The generating MCAR probability is 0.100, compared with estimates of 0.102 for the known-source fit and 0.105 for the unknown-source fit. The purpose of this example is to illustrate the two source-indicator settings, their common prediction interface, and the numerical diagnostics available for the unknown-source formulation. The numerical comparison is for one reproducible simulated sample and is not intended as a general ranking of the two formulations.

\section{Semi-synthetic application to the Blood Transfusion data}
\label{sec:realdata}

We illustrate the empirical use of \pkg{SSLfmm} with the \code{blood\_transfusion} data set~\citep{yeh2009knowledge}. The data contain \(n=748\) observations from the Blood Transfusion Service Centre classification problem. Three standardised numerical variables are used as predictors: recency, frequency, and time. The complete binary class labels are retained in \code{truth} only for evaluation, whereas \code{observed} contains the labels supplied to \pkg{SSLfmm}, with \code{NA} for observations whose labels are missing.

Because the original class labels are complete, a semi-synthetic partially labelled sample was constructed to contain both MCAR and informative MAR missingness. 10\% of the labels were removed completely at random. An additional set of relatively difficult observations, identified using disagreement among out-of-fold classifiers, was assigned to the MAR component, and the remaining labels were observed. This construction provides a partially labelled example but is not generated from the working entropy-based logistic missingness model in \pkg{SSLfmm}. Consequently, the fitted MAR parameters below should be interpreted as parameters of the working model rather than as generating parameters of the semi-synthetic construction.

The resulting fixed data set is included directly in the package, so every analysis reported below is reproducible from the distributed package object without an external data file.

\begin{Schunk}
\begin{Sinput}
R> library("SSLfmm")
\end{Sinput}
\begin{Sinput}
R> data("blood_transfusion")
R> dat <- blood_transfusion
R> table(dat$truth)
\end{Sinput}
\begin{Soutput}
  1   2
570 178
\end{Soutput}
\begin{Sinput}
R> table(dat$missing_indicator)
\end{Sinput}
\begin{Soutput}
    mar    mcar observed
     91      75      582
\end{Soutput}
\end{Schunk}

Thus, 582 observations have observed labels and 166 have missing labels, consisting of 91 MAR and 75 MCAR observations. The predictors, partially observed response, reference labels, and source indicator are prepared as follows.

\begin{Schunk}
\begin{Sinput}
R> feature_names <- c("Recency", "Frequency", "Time")
R> x <- as.matrix(dat[, feature_names, drop = FALSE])
R> y <- as.integer(dat$observed)
R> truth <- as.integer(dat$truth)
R> id_mis <- which(is.na(y))
\end{Sinput}
\end{Schunk}

The variable \code{truth} is not supplied to any fitted model; it is used only to evaluate predictions after fitting.

\subsection{Comparison of missing-label models}

We compare three missing-label formulations implemented by \code{fit\_sslfmm}(): MCAR, entropy-based MAR, and mixed. In the mixed analysis, the source of each missing label is treated as unknown. Each formulation is fitted under both equal and unequal covariance specifications, giving six fitted models.

The six analyses are fitted explicitly below so that the code block can be run from top to bottom without omitted intermediate objects.

\begin{Schunk}
\begin{Sinput}
R> fit_mcar_equal <- fit_sslfmm(x, y, g = 2, method = "mcar",
+    covariance_type = "equal", n_starts = 10, seed = 2024)
R> fit_mar_equal <- fit_sslfmm(x, y, g = 2, method = "mar",
+    covariance_type = "equal", n_starts = 10, seed = 2024)
R> fit_mixed_equal <- fit_sslfmm(x, y, g = 2, method = "mixed",
+    covariance_type = "equal", indicator = "unknown", n_starts = 10, 
+    seed = 2024)
R> fit_mcar_unequal <- fit_sslfmm(x, y, g = 2, method = "mcar",
+    covariance_type = "unequal", n_starts = 10, seed = 2024)
R> fit_mar_unequal <- fit_sslfmm(x, y, g = 2, method = "mar",
+    covariance_type = "unequal", n_starts = 10, seed = 2024)
R> fit_mixed_unequal <- fit_sslfmm(x, y, g = 2, method = "mixed",
+    covariance_type = "unequal", indicator = "unknown", n_starts = 10, 
+    seed = 2024)
\end{Sinput}
\end{Schunk}

The MCAR and MAR models use the entire partially labelled sample under a single missingness mechanism, whereas the mixed model allows MCAR and MAR components to contribute simultaneously without knowing the source of individual missing labels.

\subsection{Prediction for observations with missing labels}

All fitted models share the same prediction interface. We collect the six fits and obtain predicted classes, class probabilities, and entropy for all observations.

\begin{Schunk}
\begin{Sinput}
R> fits <- list("MCAR, equal" = fit_mcar_equal, "MAR, equal" = fit_mar_equal,
+    "Mixed, equal" = fit_mixed_equal, "MCAR, unequal" = fit_mcar_unequal,
+    "MAR, unequal" = fit_mar_unequal, "Mixed, unequal" = fit_mixed_unequal)
R> pred <- lapply(fits, predict, newdata = x, type = "all")
\end{Sinput}
\end{Schunk}

Performance is evaluated only for the 166 observations whose labels are missing during fitting. Because their feature vectors are included in the semi-supervised fit, these results should not be interpreted as out-of-sample generalisation. We report accuracy, balanced accuracy, macro F1, and log loss.

\begin{Schunk}
\begin{Sinput}
R> metric_names <- c("accuracy", "balanced_accuracy", "macro_f1", "log_loss")
R> results_missing <- matrix(NA_real_, length(pred), length(metric_names),
+    dimnames = list(names(pred), metric_names))
R> for (i in seq_along(pred)) {
+    perf <- classification_performance(truth[id_mis], pred[[i]]$class[id_mis],
+      pred[[i]]$posterior[id_mis, , drop = FALSE])
+    results_missing[i, ] <- perf$metrics[metric_names]
+ }
R> round(results_missing, 3)
\end{Sinput}
\begin{Soutput}
                       accuracy balanced_accuracy macro_f1 log_loss
MCAR, equal               0.584             0.521    0.437    0.678
MAR, equal                0.566             0.500    0.397    0.719
Mixed, equal              0.596             0.535    0.462    0.657
MCAR, unequal             0.542             0.497    0.462    1.094
MAR, unequal              0.548             0.502    0.466    1.077
Mixed, unequal            0.614             0.574    0.557    0.825
\end{Soutput}
\end{Schunk}

In this semi-synthetic example, the mixed model gives the strongest overall performance among the three missingness formulations under both covariance specifications. With equal covariance, it achieves an accuracy of 0.596, balanced accuracy of 0.535, and macro F1 of 0.462, together with the smallest log loss (0.657). Under unequal covariance, the mixed model gives the highest accuracy (0.614), balanced accuracy (0.574), and macro F1 (0.557). Its log loss is less favourable than that from the equal-covariance mixed fit, indicating a trade-off between hard classification and probabilistic performance.

\subsection{Missingness-parameter estimates}

The estimated missingness parameters provide additional information about the fitted MAR and mixed mechanisms. For the mixed model, \(\alpha\) represents the fitted MCAR contribution, while \(\xi_0\) and \(\xi_1\) describe the entropy-based MAR component.

\begin{Schunk}
\begin{Sinput}
R> missingness_estimates <- rbind(
+    "MAR, equal" = c(alpha = NA, xi0 = unname(fit_mar_equal$xi["xi0"]),
+      xi1 = unname(fit_mar_equal$xi["xi1"])),
+    "Mixed, equal" = c(alpha = fit_mixed_equal$alpha,
+      xi0 = unname(fit_mixed_equal$xi["xi0"]),
+      xi1 = unname(fit_mixed_equal$xi["xi1"])),
+    "MAR, unequal" = c(alpha = NA, xi0 = unname(fit_mar_unequal$xi["xi0"]),
+      xi1 = unname(fit_mar_unequal$xi["xi1"])),
+    "Mixed, unequal" = c(alpha = fit_mixed_unequal$alpha,
+      xi0 = unname(fit_mixed_unequal$xi["xi0"]),
+      xi1 = unname(fit_mixed_unequal$xi["xi1"])))
R> round(missingness_estimates, 3)
\end{Sinput}
\begin{Soutput}
                 alpha    xi0    xi1
MAR, equal          NA  2.596  4.997
Mixed, equal     0.086  9.343 19.802
MAR, unequal        NA -1.191  0.036
Mixed, unequal   0.109 12.369 24.785
\end{Soutput}
\end{Schunk}

The mixed fits estimate \(\alpha\) as 0.086 under equal covariance and 0.109 under unequal covariance. These values are close to the 0.10 MCAR fraction used in constructing the semi-synthetic sample. Both mixed fits also estimate a strongly positive \(\xi_1\), indicating a steep fitted increase in MAR missingness probability with classification uncertainty. The fitted slopes are large (19.802 and 24.785) and differ substantially from the MAR-only estimates. They should not be interpreted as estimates of a known generating slope, because the semi-synthetic MAR group was not generated from Equation~\ref{eq:q}. Rather, the values describe the best-fitting working missingness model for this constructed example and should be interpreted as features of that working model rather than as recovered generating parameters.

The Blood Transfusion example therefore demonstrates an applied workflow in \pkg{SSLfmm}. Alternative assumptions about label missingness are fitted through a common interface, predictions are evaluated on observations with missing labels, and the fitted mixed mechanism can be interpreted through its estimated missingness parameters. The reference labels used for these performance comparisons are available because the example is semi-synthetic; in a genuinely partially labelled application, equivalent performance measures would not generally be observable for the unlabelled cases.

\section{Final remarks}\label{sec:final}

The \pkg{SSLfmm} package provides a likelihood-based framework for semi-supervised Gaussian-mixture classification when the process governing label availability is itself part of the statistical model. Its main contribution is the mixed formulation, together with ECM algorithms for the two settings in which the source of a missing label is either known or unknown. Complete-case, MCAR, MAR, and mixed analyses are available through a common fitting interface, and the resulting objects share the same methods for inspection, prediction, and performance assessment. The examples in this paper illustrate two complementary uses of this framework. The simulation illustration shows how the known- and unknown-source formulations can be fitted to the same partially labelled sample and compared through their fitted missingness parameters and classification summaries. The semi-synthetic Blood Transfusion analysis considers a broader model comparison and demonstrates how alternative assumptions about label missingness can lead to different fitted classifiers even when the observed feature vectors are unchanged. In this constructed example, the mixed model gives the strongest hard-classification performance under both covariance specifications. These numerical results are illustrative rather than evidence that one formulation will dominate across applications. The scope of \pkg{SSLfmm} is intentionally focused. The current implementation assumes fully observed feature vectors and Gaussian class-conditional distributions, and its purpose is therefore different from that of general finite-mixture packages or software designed primarily for missing feature values. It also relies on a specified entropy-based model for informative label missingness. In particular, when the missingness source is unknown, the data may contain limited information for separating the MCAR and MAR contributions, so the fitted decomposition should be interpreted together with likelihood values, multiple starting points, and estimated missingness parameters rather than in isolation. The large slope estimates in the semi-synthetic application also warrant cautious interpretation because the working entropy-based missingness model is only an approximation to the data-construction mechanism.

Several extensions would broaden the range of problems that can be addressed within this framework. These include additional covariance parametrisation and class-conditional distributions, formal model-selection procedures for competing missingness specifications, extensions to incomplete feature vectors, and more flexible models for informative label availability. The present package provides a compact implementation for studying a point that is easily overlooked in semi-supervised classification: the unlabelled portion of a sample reflects not only unknown class memberships but also the process governing label availability, and modelling that process can be relevant to both estimation and prediction.

\section*{Acknowledgments}

This work was supported by the Australian Research Council [DP230101671].
\bibliographystyle{apalike}
\bibliography{mclachlan-wu}

\end{document}